\documentclass{aa}
\usepackage{cases}
\usepackage{graphicx}
\usepackage{txfonts}
\usepackage{longtable}
\newcommand{\e}{\epsilon}
\newcommand{\g}{\gamma}

\newcommand{\ep}{\epsilon^\prime}

\begin{document}
   \title{Constraining Extragalactic Background Light From TeV Blazars}

   \author{Jianping Yang
          \inst{1,2,3}
          \and
          Jiancheng Wang\inst{1,2}
          }

   \institute{National Astronomical Observatories, Yunnan Observatory, Chinese Academy
of Sciences,  Kunming 650011, China\\
              \email{yangjp@ynao.ac.cn}
         \and
Key Laboratory for the Structure and Evolution of Celestial Objects,
Chinese Academy of Sciences,  Kunming 650011, China
         \and
Yunnan Agricultural University, Kunming 650201, China
             }

   \date{Received 06/04/2010; accepted  17/06/2010}


  \abstract
   {}
   {Our goal is to research the upper limits on the extragalactic background light (EBL).}
   {The upper limits on the extragalactic background light (EBL), using
the Fermi and very high energy (VHE) spectra recently observed in
TeV blazars, are presented. We use an assumption that the VHE
intrinsic photon index cannot be harder than the Fermi index
measured by the Fermi-LAT.}
   {Totally, these upper limits on the EBL
are compatible with ones given by most of EBL models. However, the
models of high EBL density are denied by TeV blazars.}
 {}
\authorrunning {J. Yang \& J. Wang}
\titlerunning {Constraining Extragalactic Background Light From TeV Blazars}

   \maketitle

\section{Introduction}

The diffuse extragalactic background light (EBL) consists of the sum
of the starlight emitted by galaxies through the history of the
Universe, and includes an important contribution from the first
stars. Direct measurements of the extragalactic background light
(EBL) from the infrared (IR) to the ultraviolet (UV) are difficult
because of the light pollution of bright foreground sources (see
comprehensively reviewed measurements and implications of the cosmic
infrared background, {Hauser} \& {Dwek} 2001). The method of galaxy
counts is used to estimate the EBL, it just provides a lower limit
owing to the unknown of unresolved sources. Various models for the
EBL have been published (Salamon \& Stecker 1998; Malkan \& Stecker
1998; Malkan \& Stecker 2001; Stecker et al. 2006; Kneiske et al.
2002; Kneiske et al. 2004; Primack et al. 2005; Primack et al. 2008;
Gilmore et al. 2008; Gilmore et al. 2009; Franceschini et al. 2008;
Razzaque et al. 2009; Finke et al. 2010). These models include
different degrees of complexity, observational constraints and data
inputs.

Absorption features imprinted on the very high energy (VHE) spectra
of distant extragalactic objects by background light photons provide
an indirect approach to study the EBL. Assuming an intrinsic
gamma-ray spectrum, \cite{stecker93,stanev98} have constrained the
EBL from the observed VHE spectra of blazars. \cite{aharonian06}
have also discussed upper limits on the background light at
optical/near-infrared wavelengths based on the HESS observation of
1ES 1101-232. They assume the intrinsic spectrum to be not harder
than $\Gamma_{int} = 1.5$ and put limits on EBL quite close to the
lower limits by galaxy counts. The detail studies of EBL shapes and
blazar VHE spectra are also given by
\cite{mazin07,schroedter05_EBL,Finke09}, where same intrinsic
spectrum for all blazars is assumed. In fact, blazars have different
intrinsic spectra. The main handicap of this approach to limit EBL
is the uncertainty about the intrinsic spectrum of VHE.

To date, 35 AGN sources have been detected at TeV energies(E$>$100
GeV)\footnote[1]{update see:
http://www.mppmu.mpg.de/\~{}rwagner/sources/}. Their observed VHE
spectra have power law shapes with the index $\Gamma_{VHE} \geq $ 2,
in which distant sources have large $\Gamma_{VHE}$, up to 4 (e.g.
Acciari et al. 2009b; Albert et al. 2007b, 2008b; Aharonian et al.
2006c, 2005a). Many of these sources have recently been detected at
GeV energies by the Fermi Gamma-ray Space Telescope(Abdo et al.
2009; 2010b). \cite{Abdo09_TeV} have extrapolated the Fermi spectrum
up to 10 TeV assuming a single spectral index and taken it as an
intrinsic spectrum of VHE ranges. The most break of the observed VHE
spectra are consistent with the absorption predicted by the minimal
EBL density model. For a TeV source, the presence of a break between
the Fermi and VHE energy range might be caused by some internal or
external factors. The internal factors include a break of emitting
particle distribution or an intrinsic absorption caused by strong
optical-infrared radiation within the source (Donea \& Protheroe
2003). The external factors usually refer to the cosmic attenuation
effect. Furthermore, it is difficult to well predict the intrinsic
spectrum from simultaneous multi-wavelength observations because of
the complexity of VHE emission mechanism. In this work, we assume
that the Fermi spectral index measured by Fermi-LAT is the lower
limit of intrinsic VHE spectral index for TeV blazars instead of
single $1.5$. In the other words, the photon index from the Fermi to
VHE energy range is only softened except the presence of a new
component (Yang \& Wang 2010), or monochromatic radiation fields
within the source (Aharonian et al 2009a). We note that the steeper
intrinsic index assumed by us provides stronger constraints to the
EBL intensity (all Fermi photon index of TeV sources larger than the
1.5 with the exclusion of H 1426+428 $\Gamma_{Fer}$ = 1.47).
Moreover, taking into account the differences of VHE emission
between the different sources, the assumption is more reasonable
than one of $\Gamma_{int}>1.5$. Recently \cite{Georganopoulos10} and
\cite{Mankuzhiyil10} also use the extrapolation of the Fermi data as
upper limits of intrinsic TeV spectra. Assuming the VHE intrinsic
spectra corrected by EBL absorption be softer than the Fermi
spectra, \cite{Prandini10} found that the derived redshifts are
larger than true ones. It shows that their assumption is reliable.
Based on this assumption, we analyze the Fermi and VHE spectra of
TeV blazars and give upper limits on the EBL intensity.

In \S\ 2 we describe the method of calculating $\gamma - \gamma$
absorption optical depth, $\tau_{\gamma \gamma}(\epsilon)$, and the
EBL intensity, developed by \cite{schroedter05_EBL} and
\cite{Finke09}. In \S\ 3 we apply the method to the TeV blazars with
VHE and Fermi spectra and discuss the limits of these sources on the
EBL.

\section{THE METHOD}

The VHE absorption of the EBL is caused by the pair production of
photon-photon collision. The observed VHE flux is given by
\begin{eqnarray}
f_{obs}(E_{\gamma}) = e^{-\tau(E_{\gamma})}f_{int}(E_{\gamma})\ ,
\end{eqnarray}
where $\tau(E_{\gamma})$ is the optical depth, $E_{\gamma}$ is the
observed $\g$-ray photon energy, and $f_{int}(E_{\gamma})$ is the
intrinsic flux.

For a VHE source at redshift $z_e$, the optical depth of its
$E_\gamma$ energy photon caused by the EBL is given by
\begin{eqnarray} \label{tau_detail}
\tau(E_\gamma,z_e)  = c \pi r_e^2 (\frac{m^2 c^4}{E_{\gamma}})^2
\int_0^{z_e} dz {dt \over dz }
   \int_{\frac{ m^2c^4}{E_\gamma  (1+z)}}^\infty d\epsilon \cdot
 \epsilon^{-2} n_(\epsilon, z) \bar{\varphi}[s_0(\epsilon)],
 \end{eqnarray}

where $n (\epsilon, z)$ is the photon number density of the EBL with
energy $\epsilon$ at redshift $z$, $r_e$ is the classical electron
radius, $s_0=\epsilon E_{\gamma}/m^2c^4$,
 $\bar{\varphi}[s_0(\epsilon)]$ is a function given by
 \cite{gould67}, and $\frac{dt}{dz}$ is the differential time of
 redshift given by
\begin{eqnarray}
 dt/dz = \frac{1}{H_0 (1+z)} \left[(1+z)^2 (1+\Omega_m z) -
z(z+2)\Omega_\Lambda \right]^{-1/2} .
 \end{eqnarray}

\cite{Abdo09_TeV} have found that the intrinsic spectra of many TeV
sources can be described by a single power-law across the Fermi and
VHE energy ranges. In fact, the observed Fermi and VHE spectral
indices are different duo to the EBL absorption. Their difference
$\Delta \Gamma$ increases with redshift. For example, M 87 and Cen A
with low redshidts have $\Delta \Gamma \approx $ 0, while blazars
with redshifts greater than 0.1 show $\Delta \Gamma \geq $ 1.5. We
assume that the observed Fermi spectral index is the lower limit of
intrinsic VHE spectral index, i.e., $\Gamma_{int,VHE}^{min} \approx
\Gamma_{Fer}$. For some objects, multiple VHE spectra have been
observed in different flux states. We adopt the VHE spectra of their
low flux states to constrain the EBL.

Based on $\Gamma_{int,VHE}^{min}$, an upper limit on the optical
depth $\tau(E_{\g},z_e)$ is given by \cite{Finke09}
\begin{eqnarray}
\label{taumax} \tau^{max}(E_{\g},z_e) =
    \tau(E_{\g,min},z_e) +
    (\Gamma_{obs}-\Gamma_{int,VHE}^{min})\ln(E_{\g}/E_{\g,min}),\
 \end{eqnarray}
and its standard error $\tau^{max}$ is given by
\begin{eqnarray}
\label{sigmataumax} \sigma(\tau_{\g\g}^{max}) =
\sigma(f_{obs}(E_{\g})) / f_{obs}(E_{\g}),\
\end{eqnarray}
where $\tau(E_{\g,min},z_e)$ at the lowest energy $E_{\g,min}$ of
VHE observations is estimated by the EBL model of
\cite{franceschini08}.

Now we use $\tau^{max}(E_\gamma,z_e)$ to estimate an upper limit on
the EBL number density. Following \cite{schroedter05_EBL,Finke09},
we take $dt/dz \approx H_0^{-1}$ for TeV sources due to low
redshift, where $H_0 = 70$ km s$^{-1}$ Mpc$^{-1}$. We assume the
monochromatic absorption of VHE photon $E_{\g}$ by the EBL at the
energy $\ep=2m^2c^4/(E_{\g}(1+z))\approx 2m^2c^4/E_{\g}$ where the
pair-production cross section reaches the largest value, and give an
upper limit on the EBL number density, $n(\e,z)$. Using the Dirac
delta-function, we approximately write $n(\e,z)$ as
\begin{eqnarray}
n(\e,z\approx0 ) \approx \ep n(\ep,z\approx0) \delta (\e-\ep).
\end{eqnarray}
Integrating Eq.(\ref{tau_detail}), we obtain:
\begin{eqnarray}
\label{nEBLmax} n(\ep,z\approx0) = \frac{2H_0 \tau^{max}_{\g\g}
E_{\gamma}}
    {c z_e \pi r_e^2 m^2c^4 \bar{\varphi}(2)},
\end{eqnarray}
where $\bar{\varphi}(2) \approx 1.787$.  The error of the EBL number
density is given by
\begin{eqnarray}
\sigma(n) =\frac{2H_0 \sigma(\tau_{\g\g}^{max}) E_{\gamma}}
    {c z_e \pi r_e^2 m^2c^4 \bar{\varphi}(2)}.
\end{eqnarray}
Finally the EBL intensity is given by
\begin{eqnarray}
\nu I_\nu(z) = \frac{c}{4\pi}\ \e ^2 n(\e,z)\ .
\end{eqnarray}

\section{Results and Discussion}

The EBL has two spectral humps with different origins. The blue hump
at UV-Optical-NIR (near-infrared) wavelengths comes from stars. The
red hump at MIR (mid-infrared) and FIR (far-infrared) wavelengths is
from the absorption and re-emission of starlight by the interstellar
medium. Therefore, the EBL includes the important information of
star formation and evolution.

\begin{figure}
\includegraphics[width=100mm]{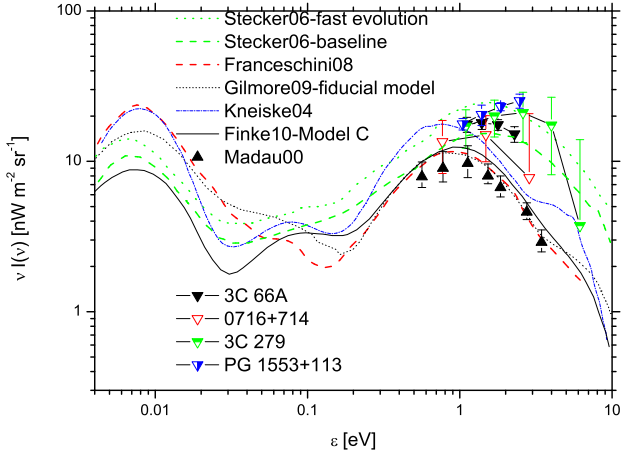}
\caption{Upper limits of the EBL given by 3C 66A, 0716+714, 3C 279
and PG 1553+113 under the assumption that the Fermi spectral index
measured by Fermi-LAT can be used as an lower limit of the intrinsic
VHE spectral index. Also plotted are several EBL models: the
baseline and fast evolution models of \cite{stecker06a}(dash, dot
green curves, respectively), the model of \cite{franceschini08}(dash
red curve), the fiducial model from \cite{gilmore09}(dot black
curve), the best fit model from \cite{kneiske04}(dot-dash blue
curve), and the Model C of \cite{Finke10}(solid black curve). We
also list the lower limits of EBL from source counts with upward
triangles (Madau \& Pozzetti 2000). \label{fig.1}}
\end{figure}

The TeV blazars used to constrain the EBL are listed in Table 1,
where the spectra at low-flux state are used to the utmost, since
the 11 months averaged Fermi spectra are unlikely to correspond to
the high state. Through calculation, we find that four blazars, 3C
66A, 0716+714, 3C 279, and PG 1553+113, give stronger constraint on
the EBL density. Since other TeV blazars are consistent with all
listed EBL models within the error range, we do not give their
constraint. The EBL upper limits given by the spectra of four TeV
blazars are shown in Fig. \ref{fig.1}. The curves of several EBL
models are also plotted in Fig. \ref{fig.1}:
\cite{kneiske04,gilmore09,stecker06a,Finke10,franceschini08}. In
this work the calculated limits on the EBL at UV-Optical-NIR
wavelengths are strong. For comparison, we also list the lower
limits of the EBL from source counts \cite{Madau00}, shown by the
upward triangles. These calculated limits are inconsistent with the
fast evolution model given by \cite{stecker06a} in NIR-Optical
wavelengths, but are still compatible with their baseline model. For
the fast evolution model of \cite{stecker06a} the extinction of UV
photons by the interstellar gas in galaxies is not considered, the
UV and Optical-NIR photon density might be over-estimated. In fact,
the observed gamma-ray hard spectra of H 2356-309 (z = 0.165) and
1ES 1101-232 (z = 0.186) by \cite{aharonian06} suggest that an upper
limit to the EBL at optical-NIR wavelengths is very close to the
lower limit given by the integrated light of resolved galaxies. This
implies that the EBL is more transparent to high energy
$\gamma$-rays than previously thought and the contribution from
sources except starlight is less. \cite{Essey09} suggest a new
interpretation of these observations. For distant blazars, the
gamma-ray emission is dominated by the secondary photons, while for
nearby blazars the emission is from the primary photons. Therefore,
they argue that distant AGN would show no significant attenuation
due to pair production on the EBL. Also, we should note that the
redshift of 3C 66A, assumed to be z = 0.444, has large uncertain
(Miller et al. 1978). If its redshift is less than 0.444, the upper
limits obtained will be relaxed. Due to the very bright nucleus, the
redshift of 0716+714 is still uncertain. Stickel et al. (1993)
repeated spectroscopic observations without obtaining the redshift,
but they found that two neighboring galaxies have quite similar
redshifts of 0.26. However, its host galaxy detection gave the
redshift of $z=0.31$ (Nilsson et al. 2008). The redshift of PG
1553+113 also has large uncertain. \cite{Sbarufatti05} derived its
lower limit of z$>$0.78. \cite{Sbarufatti06} used the spectra of ESO
VLT to give a limit of z$>$0.09. \cite{Abdo10} constrained z $\leq $
0.75 combining Fermi and VHE gamma-ray data. \cite{Danforth10}
constrained its redshift to be z $\sim $ 0.4 -- 0.6 by Hubble Space
Telescope. In this work, we adopt its redshift as z=0.78. While 3C
279 has well-known redshift. Totally, in spite of large redshift
uncertainty for some objects, the models of high EBL density are
denied by 3C 279.

\begin{table*}

\begin{minipage}{140mm}
\caption{TeV Blazar Sample} \centering
\begin{tabular}{@{}lcccccc}

 \hline

 Blazar  &  Redshift  &  Fermi photon Index ($\Gamma_{Fer}$)   &
VHE photon Index($\Gamma_{VHE}$)  &  $E_{min}$ [TeV]  &  $E_{max}$
[TeV]  &  Reference \\
 \hline
3C 66A        & 0.444 & $1.93 \pm 0.02$ &  $4.1  \pm 0.4 $ & 0.23 & 0.47   &1 \\
S5 0716+714   & 0.31   & $2.15 \pm 0.03$ &  $3.45 \pm 0.54$ & 0.18 & 0.68   &2  \\
1ES 0806+524  & 0.138 & $2.1 \pm 0.1$ &  $3.6  \pm 1.0 $ & 0.31 & 0.63   &3 \\
1ES 1011+496  & 0.212 & $1.93 \pm 0.04$ &  $4.0  \pm 0.5 $ & 0.15 & 0.59   &4  \\
Mark 421      & 0.031 & $1.81 \pm 0.02$ &  $2.2  \pm 0.08$ & 0.13 & 2.86   &5 \\
Mark 180      & 0.046 & $1.86 \pm 0.11$ &  $3.3 \pm 0.7  $ & 0.18 & 1.32   &6  \\
1ES 1218+304  & 0.182 & $1.7 \pm 0.08$ &  $3.08 \pm 0.34$ & 0.19 & 1.44   &7 \\
W   Comae     & 0.102 & $2.06 \pm 0.04$ &  $3.81 \pm 0.35$ & 0.27 & 1.15   &8  \\
3C 279        & 0.536 & $2.32 \pm 0.02$ &  $4.11 \pm 0.68$ & 0.08 & 0.47   &9 \\
H   1426+428  & 0.129 & $1.49 \pm 0.18$ &  $3.5 \pm 0.35 $ & 0.82 & 5.66   &10  \\
PG 1553+113   & 0.78  & $1.66 \pm 0.03$ &  $4.0  \pm 0.6 $ & 0.21 & 0.50   &11 \\
Mark 501      & 0.034 & $1.85 \pm 0.04$ &  $2.45 \pm 0.07$ & 0.14 & 4.58   &12  \\
1ES 1959+650  & 0.048 & $2.1 \pm 0.05$ &  $2.58 \pm 0.18$ & 0.19 & 2.40   &13 \\
PKS 2005-489  & 0.071 & $1.9 \pm 0.06$ &  $4.0  \pm 0.4 $ & 0.23 & 2.27   &14  \\
PKS 2155-304  & 0.117 & $1.91 \pm 0.02$ &  $3.32 \pm 0.06$ & 0.23 & 2.27   &15 \\
BL  Lacertae  & 0.069 & $2.38 \pm 0.04$ &  $3.6  \pm 0.5 $ & 0.16 & 0.70   &16  \\
1ES 2344+514  & 0.044 & $1.57 \pm 0.12$ &  $2.95 \pm 0.12$ & 0.19 & 4.00   &17 \\
\hline


\label{blazartable} \\
\end{tabular}
\end{minipage}
\\
The Fermi data come from the \cite{Abdo10_1LAC}. VHE data refer to:
(1)\cite{acciari09_3c66a}; (2) \cite{Anderhub09_0716}; (3)
\cite{acciari09_0806}; (4) \cite{albert07_1011}; (5)
\cite{albert07_421}; (6) \cite{albert06_mrk180}; (7)
\cite{acciari09_1218}; (8) \cite{acciari08_wcomae}; (9)
\cite{albert08_279}; (10) \cite{aharonian03_1426}; (11)
\cite{aharonian06_1553}; (12) \cite{albert07_501}; (13)
\cite{Tagliaferri08_1959}; (14) \cite{aharonian05_2005}; (15)
\cite{aharonian05_2155}; (16) \cite{albert07_bllac}; (17)
\cite{albert07_2344} . \\
\end{table*}


In low redshift, we assume the EBL do not evolve with redshift.
However, the EBL is progressively generated by galaxies and active
nuclei (AGN) during most of the Hubble time, particularly below z =
1. The evolution of their photon number density is a very complex
function of time and frequency.

\begin{figure}
\includegraphics[width=100mm]{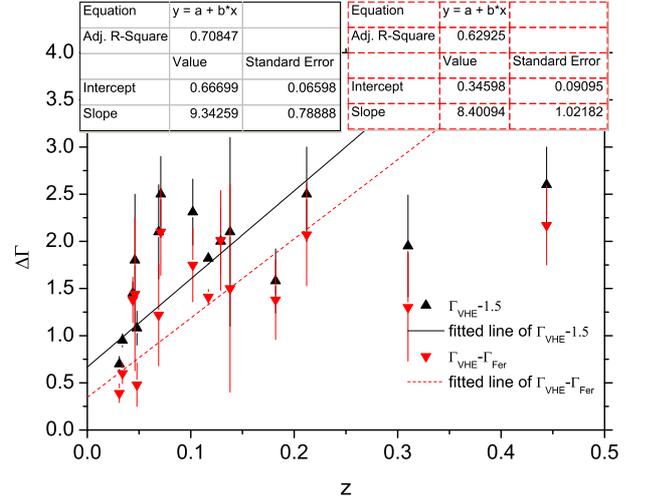}
\caption{Difference, $\Delta \Gamma$, between the measured VHE and
Fermi photon indices (or conventional limit 1.5) as a function of
the redshift. Red inverted triangles denote the
$\Gamma_{VHE}-\Gamma_{Fer}$, and black triangles denote the
$\Gamma_{VHE}-1.5$. The red dash line show the fitting of
$\Gamma_{VHE}-\Gamma_{Fer}$.\label{fig.7}}
\end{figure}

The main problem of TeV blazars observations limiting on the EBL is
the uncertainty of VHE intrinsic spectra. In the early works, all
TeV blazars are assumed to have a lower limit of $\Gamma_{int}=1.5$
based on some theories of particles acceleration and emission (e.g.
Aharonian et al 2006a). The same assumption was also adopted by
\cite{schroedter05_EBL} with the analysis of observed correlation
between their $\Gamma_{VHE}$ and redshift. \cite{stecker06b} have
derived a simple analytic approximation of the EBL absorption on the
spectra of TeV blazars in the energy ranges of 0.2 TeV $<$ E $<$ 2
TeV, and found that $\Delta \Gamma(z) = \Gamma_{VHE} - \Gamma_{int}$
is a linear function of the redshift z in the range of 0.05 - 0.4.
In this work, we assume the Fermi spectral index measured by
Fermi-LAT to be the lower limit of intrinsic VHE spectral index. In
Fig. \ref{fig.7}, we compare the redshift variation of
$\Gamma_{VHE}-\Gamma_{Fer}$ and $\Gamma_{VHE}-1.5$ in the range of
$0.03 < z < 0.4$ excluding 3C 66A (z=0.444), 3C 279 (z=0.536) and PG
1553+113 (z=0.78). It is shown that they have similar correlation
with redshift, but the parameters fitting
$\Gamma_{VHE}-\Gamma_{Fer}$ are more similar with C and D of the
baseline model given by \cite{stecker06b} and \cite{Stecker10}. In
fact, the baseline model gives lower EBL density than the fast
evolution model does, high EBL density models are disfavored by
observations, such as \cite{aharonian06}; \cite{Georganopoulos10}.
Most of the $\Gamma_{VHE}-\Gamma_{Fer}$ are less than the
$\Gamma_{VHE}-1.5$, and smaller $\Delta \Gamma$ will provide
stronger constraints to EBL. If using $\Gamma_{int} \geq 1.5$ to
limit the EBL is reasonable, $\Gamma_{int} \geq \Gamma_{Fer}$ will
be more feasible. 3C 66A and 0716+714 obviously deviates the
correlation shown in Fig. \ref{fig.7}, it implies that the assumed
redshift might be uncorrect.

Since the simultaneous data of Fermi and VHE are less available
nowadays, we only use non-simultaneous spectra. We also note that
the upper limits of EBL density are great depended on the VHE photon
index (see the equation (\ref{taumax})). In fact, no significant
spectral variability is observed in VHE bands. For example, PG
1553+113 has many times VHE observations (Aharonian et al. 2006b,
Aharonian et al. 2008, Albert et al. 2007f, Albert et al. 2009), its
photon index is very similar as observed by HESS and MAGIC (Abdo et
al. 2010a). For PKS 2155-304, no significant spectral variability
appears despite flux variation with a factor of two (Aharonian et
al. 2009). For 1ES 1218+304, \cite{acciari09_1218flare} found that
the VHE spectral shape has no change between flare and quiescent
state. Therefore our calculated results are meaningful when VHE
spectra have no obvious variation.

We use the assumption of monochromatic absorption to calculate the
EBL intensity at specified wavelengths. The result will be larger
than the actual case due to the EBL photons near specified
wavelength also contributing the absorption. But, this method
proposed by \cite{schroedter05_EBL,Finke09} does not assume the EBL
spectrum and has the advantage compared to other methods because the
EBL spectrum is not easily known. Future observations of Fermi and
VHE spectra for blazars will provide strong constraints on the EBL.

\section*{Acknowledgments}

We thank the referee for a very helpful and constructive report that
helped to improve our manuscript substantially. We acknowledge the
financial supports from the National Natural Science Foundation of
China 10673028 and 10778702, the National Basic Research Program of
China (973 Program 2009CB824800) and the Policy Research Program of
Chinese Academy of Sciences (KJCX2-YW-T24).

\clearpage

\end{document}